\documentclass[10pt,a4paper,onecolumn]{article}
\usepackage{graphicx}
\usepackage{color}
\begin{document}
\title{Efficiency at the maximum power of the power law dissipative
Carnot-like Heat engines with non-adiabatic dissipation}

\author{M. Ponmurugan \\
Department of Physics, School of Basic and Applied Sciences, \\
Central University of Tamilnadu, Thiruvarur - 610 005,  \\
Tamilnadu, India. e-mail:ponphy@cutn.ac.in}



\maketitle

\begin{abstract}
We study the efficiency at the maximum power of non-adiabatic dissipative 
(internally dissipative friction in finite time adiabatic processes) Carnot-like heat engines
operate in finite time under the power law dissipation regime.  
We find that the non-adiabatic dissipation does not influence the universal
minimum  and maximum  bounds on the efficiency at the maximum power obtained 
in the generalized  dissipative Carnot-like heat engines 
which does not take in to account the non-adiabatic dissipation.  
\end{abstract}


\section{Introduction}

Finite time thermodynamic studies of heat engines operate between hot and cold heat reservoirs focus mainly on improving its performance towards the traditional equilibrium Carnot engine \cite{book1}. Carnot engine follows a particular cycle of infinite long duration called as the Carnot cycle, which consists of two isothermal 
and two adiabatic processes. The engine efficiency is defined as $\eta=W/Q_h$, where $W$ is the work performed and 
$Q_h$ is the amount of heat absorbed from the hot reservoir at a higher temperature $T_h$, while the engine 
delivers the heat $Q_c$ to the cold reservoir at temperature $T_c$. The efficiency of  finite time 
heat engines are bounded below the Carnot engine efficiency, $\eta_C=1-T_c/T_h$. The power delivered by the 
cyclic heat engine is $P=W/\tau$, where $\tau$ is the total time taken to complete the given cycle.

There is a trade off between power and efficiency. The general belief is that $\eta_C$ can be achieved only at zero power. This has been proved for different systems working under different conditions, such as, thermoelectric classical and quantum systems \cite{brandf1,brandf2}, micro and nano systems driven by periodic temperature variations \cite{brandf3}, periodically driven system based on Onsager Coefficient \cite{proes},  thermoelectric transport of quantum system \cite{yam} and  master equation approach of classical 
and quantum systems \cite{shi1,shi2}.  While several other studies showed that $\eta_C$ also  
attained at nonzero power.  Few of the studies for the thermoelectric system with broken time-reversal symmetry \cite{ben,brand}, classical and quantum Carnot cycle at finite power \cite{alla,holu},
efficiency statistics of heat devises \cite{pole}, quantum Otto engine with finite-size scaling \cite{camp} and quantum Otto cycle of two-level system \cite{leg}. Studies based on sub-linear transport 
law \cite{koe},  stochastic thermodynamics \cite{pol,lee}, steady state heat engine \cite{steady}, electrically charged black hole \cite{john}, classical harmonic oscillator under 
linear response regime \cite{bona} and  trapped Bose gas in a quantum heat engine \cite{trap} showed that 
the system efficiency approaches $\eta_C$ at finite power. Recent studies based on classical Markov 
processes \cite{shi3,shi4}, fluctuation of work and power \cite{holu1,holuuprl}, exergy \cite{pon1},
quantum dot model with zero entropy production \cite{lee1} and information engines \cite{bau,infor}
also proved the attainability of $\eta_C$ at nonzero power.

The real heat engines operate in the finite time duration of non-zero power output with the efficiencies  less than $\eta_C$. Different optimization procedures are introduced to enhance the performance of heat engines  \cite{targetfn1,targetfn2,targetfn3,multitarget}. In particular, the efficiency at maximum power, $\eta_P$, is being used frequently to investigate the improved  performance of heat engines. This can be done by optimizing the efficiency of heat engines at maximum power condition. Earlier attempts to optimize the heat engine efficiency at maximum power was investigated by  different researchers independently by employing different formulations, which  are generally called as the Curzon-Ahlborn efficiency and is given by $\eta_{CA}=1-\sqrt\frac{T_c}{T_h}=1-\sqrt{1-\eta_C}$ \cite{novikov,curzon}. The efficiency at maximum power obtained from this formulation closely matches with the observed efficiencies of real heat engines \cite{geneff}. Nevertheless, these models does not provide the universal bounds on the efficiency at maximum power \cite{book1,Hoffmann}, which is the central focus of the present work.

The phenomenological model of the finite time heat engine proposed to study the universal bounds on the efficient at maximum power is  the low-dissipation Carnot engine \cite{Espo,lowdiffcycle}. The study on the microscopic model of quantum heat engine indicated  the presence of the power law like dissipation of Carnot-like heat engines \cite{cavina}. The generalization of the low-dissipation  model was discussed earlier \cite{yang} and
also showed recently  that \cite{ponlow} the power law dissipation incorporated in the model provides the generalized universal nature of lower bound, $\frac{\eta_C}{(\delta+1)}$, and  the upper bound
$\frac{\eta_C}{(\delta+1)-\delta \eta_C}$ on the efficiency at maximum power,
where $\delta \ge 0$ is the degree of power law dissipation. 
The above results are derived under the assumption that the time taken to complete the adiabatic expansion and compression processes are negligible.

There are few studies on the low-dissipation model, which include the non-adiabatic dissipation in finite time adiabatic processes \cite{adiaheat,adiaheatqm,adiarefri}. The dissipation due to the effects of inner friction during the finite time adiabatic process is known as non-adiabatic dissipation \cite{adiaheatqm,infriction}.
These studies showed that the non-adiabatic dissipative term additionally incorporated  in the low-dissipation model does not influence the extreme bounds on the efficiency at maximum power \cite{adiaheat,adiaheatqm}. This raises a question whether such a non-adiabatic dissipative term can influence the universal bounds on the efficiency at maximum power of the power law dissipative Carnot-like heat engines? In order to answer this, in this paper, we study the efficiency at maximum power of non-adiabatic dissipative Carnot-like heat engines operate in a finite time under the power law dissipation regime \cite{yang,ponlow}.

\section {Power law dissipation Carnot-like heat engine}
The power law dissipation Carnot-like heat engine model follows the cycle consisting  of two isothermal processes of finite time duration and two instantaneous adiabatic processes. The  working substance is in contact with the hot  reservoir at temperature $T_h$ in the isothermal expansion and at temperature $T_c$ in the isothermal compression during the time interval  $t_h$ and $t_c$, respectively. The amount of heat $Q_h$ and $Q_c$ exchanged between the hot
and cold reservoirs and the working substance are modeled as \cite{Espo,yang,ponlow}
\begin{eqnarray}\label{powqh}
 Q_h=T_h \left\{ \Delta S -  \Delta S_h^{ir} \right\},  
\end{eqnarray}
\begin{eqnarray}\label{powqc}
 Q_c=T_c \left\{-\Delta S - \Delta S_c^{ir} \right\},  
\end{eqnarray}
where $\Delta S_i^{ir}=\alpha_i \left(\frac{\sigma_i}{t_i}\right)^{\frac{1}{\delta}}$, $i : h, c$, are 
the irreversible entropy production, $\sigma_i=\lambda_i\Sigma_i$, in which $\Sigma_i$ are the isothermal dissipation coefficients, $\lambda_i$ \& $\alpha_i$ are  the tuning parameters and $\pm \Delta S$ is the change in entropy of the working substance during  isothermal expansion (+) and compression (-), which ensures that the system has zero change in total entropy for the cyclic process \cite{Espo}. 
The $\delta \ge 0 $ is a  real number that represents the level of (power law) dissipation present 
in the system \cite{ponlow}. The system is in normal or low-dissipation regime
with $\delta =1$, when it deviates below ($0<\delta<1$) and  
above ($\delta > 1$) from the low-dissipation regime  
are called, respectively, sub and super dissipation regime \cite{yang}.
Since the efficiency obtained by the practical heat engines are not necessarily in
the low-dissipation regime \cite{geneff}, our earlier results showed that 
the heat engines might also operate in the sub or super dissipation regime \cite{ponlow}.

The recent experimental study validated  $1/\tau$ scaling of the 
irreversible entropy production in a finite-time isothermal process
when the system undergoes a long contact time $\tau$ with the thermal bath \cite{mascal}.
Further, a recent theoretical study on quantum Otto engine showed (in terms of extra adiabatic work)
$1/\tau^2$ scaling of the irreversible entropy production in a finite-time adiabatic process \cite{boost}.
Here $\tau$ is the controlling time of the long-time adiabatic process. The authors showed that the 
special control schemes of the finite-time adiabatic process improved the maximum power and 
the efficiency  of the finite-time Otto engine.

In the case of including the non-adiabatic dissipation in our model, $Q_c$ contains additionally the 
irreversible entropy production $\Delta S_a^{ir}$ and $\Delta S_b^{ir}$, respectively, 
during the finite time adiabatic expansion and compression \cite{adiaheat,adiaheatqm,adiarefri}. 
These entropy productions can also be assumed to obey 
the power law dissipation with respect to time for completing the adiabatic processes, which are 
of the form, $\Delta S_j^{ir}=\alpha_j \left(\frac{\sigma_j}{t_j}\right)^{\frac{1}{\delta}}$, $j : a, b$,
where $\sigma_j=\lambda_j\Sigma_j$, in which $\Sigma_j$ are the adiabatic dissipation 
coefficients \cite{adiaheat}, $\lambda_j$ \& $\alpha_j$ are the corresponding tuning 
parameters in the adiabatic process.

The amount of heat $Q_c$ exchanged between the cold reservoir and the working substance which contains 
the entropy productions due to non-adiabatic dissipations is given by \cite{adiaheat,adiaheatqm,adiarefri}
\begin{eqnarray}\label{adiaqc}
 Q_c&=&T_c \left\{-\Delta S - \Delta S_c^{ir} - \Delta S_a^{ir} - \Delta S_b^{ir} \right\},  \\ \nonumber
     &=& T_c \left\{-\Delta S - \sum_{j=c,a,b}{\Delta S_j^{ir}} \right\} = T_c \left\{-\Delta S - \sum_{j=c,a,b}{\alpha_j \left(\frac{\sigma_j}{t_j}\right)^{\frac{1}{\delta}}} \right\}.  
\end{eqnarray}
Work performed by the engine during  the  total time period $\tau=t_h+t_c+t_a+t_b$ is $-W=Q_h+Q_c$. The convention used here is that the work and heat absorbed by the system are positive \cite{Espo}. 
The power generated during the Carnot cycle is, 
\begin{eqnarray}\label{pow}
  P =\frac{-W}{\tau}=\frac{1}{\tau} \left\{(T_h-T_c) \Delta S - T_h {\alpha_h \left(\frac{\sigma_h}{t_h}\right)^{\frac{1}{\delta}}} - T_c \sum_{j=,c,a,b} {\alpha_j \left(\frac{\sigma_j}{t_j}\right)^{\frac{1}{\delta}}}\right\}  . 
\end{eqnarray}
Using Eq.(\ref{powqh}), $Q_c$ can be rewritten as 
 \begin{eqnarray}\label{gqcmod}
  Q_c=-T_c \left\{\frac{Q_h}{T_h}+\sum_{i=h,c,a,b}{\alpha_i\left(\frac{\sigma_i}{t_i}\right)^\frac{1}{\delta}}\right\}.  
\end{eqnarray}
Then,
\begin{eqnarray}\label{gqcomb}
  Q_h+Q_c = \eta_C Q_h- T_c \left\{\sum_{i=h,c,a,b}{\alpha_i\left(\frac{\sigma_i}{t_i}\right)^\frac{1}{\delta}}
	 \right\}.
\end{eqnarray}
The engine efficiency during the Carnot cycle is  
 \begin{eqnarray}\label{geffg}
  \eta =\frac{Q_h+Q_c}{Q_h}=\eta_C - \frac{T_c}{Q_h} \left\{\sum_{i=h,c,a,b}{\alpha_i\left(\frac{\sigma_i}{t_i}\right)^\frac{1}{\delta}}\right\}.
	\end{eqnarray}
Using Eq.(\ref{powqh}), the above equation can be rewritten as, 
\begin{eqnarray}\label{geffg1}
  \eta &=& \eta_C -\frac{T_c}{T_h \left[\frac{\Delta S}{\alpha_h} \left(\frac{t_h}{\sigma_h}\right)^\frac{1}{\delta} -1 \right]} \left\{1+ \sum_{j=c,a,b}{\frac{\alpha_j}{\alpha_h}\left(\frac{\sigma_j t_h}{\sigma_h t_j}\right)^\frac{1}{\delta}}\right\}  
\end{eqnarray}
and the power generated during the Carnot cycle is given by 
\begin{eqnarray}\label{gpow}
  P &=& \frac{1}{\tau}\left\{(T_h-T_c) \Delta S - T_h \alpha_h\left(\frac{\sigma_h}{t_h}\right)^\frac{1}{\delta} - T_c
	\sum_{j=c,a,b}{\alpha_j\left(\frac{\sigma_j}{t_j}\right)^\frac{1}{\delta}} \right\} .
\end{eqnarray}
The values of $t_i (i:h,c,a,b)$ at which the power becomes maximum are given by,
\begin{eqnarray}\label{gtopt}
  t_h &=&\left \{ \frac{\alpha_hT_h\sigma_h^\frac{1}{\delta}}{(T_h-T_c) \Delta S} \left(1+\frac{1}{\delta}\right) \left[1+\sum_{j=c,a,b}{\left(\frac{\alpha_j T_c}{\alpha_h T_h}\right)^\frac{\delta}{\delta+1}\left(\frac{ \sigma_j}{\sigma_h}\right)^\frac{1}{\delta+1}} \right] \right \}^{\delta}, \\  \nonumber
	t_c &=& \left \{\frac{\alpha_c T_c\sigma_c^\frac{1}{\delta}}{(T_h-T_c) \Delta S} \left(1+\frac{1}{\delta}\right)\left[1+\left(\frac{\alpha_h T_h}{\alpha_c T_c}\right)^\frac{\delta}{\delta+1}\left(\frac{\sigma_h}{\sigma_c}\right)^\frac{1}{\delta+1}+\sum_{j=a,b}{\left(\frac{\alpha_j}{\alpha_c}\right)^\frac{\delta}{\delta+1}\left(\frac{\sigma_j}{\sigma_c}\right)^\frac{1}{\delta+1}} \right] \right \}^{\delta}, \\ \nonumber
	t_a &=& \left \{\frac{\alpha_a T_c\sigma_a^\frac{1}{\delta}}{(T_h-T_c) \Delta S} \left(1+\frac{1}{\delta}\right)\left[1+\left(\frac{\alpha_h T_h}{\alpha_a T_c}\right)^\frac{\delta}{\delta+1}\left(\frac{\sigma_h}{\sigma_a}\right)^\frac{1}{\delta+1}+\sum_{j=c,b}{\left(\frac{\alpha_j}{\alpha_a}\right)^\frac{\delta}{\delta+1}\left(\frac{\sigma_j}{\sigma_a}\right)^\frac{1}{\delta+1}} \right] \right \}^{\delta}, \\  \nonumber
	t_b &=& \left \{\frac{\alpha_b T_c\sigma_b^\frac{1}{\delta}}{(T_h-T_c) \Delta S} \left(1+\frac{1}{\delta}\right)\left[1+\left(\frac{\alpha_h T_h}{\alpha_b T_c}\right)^\frac{\delta}{\delta+1}\left(\frac{\sigma_h}{\sigma_b}\right)^\frac{1}{\delta+1}+\sum_{j=c,a}{\left(\frac{\alpha_j}{\alpha_b}\right)^\frac{\delta}{\delta+1}\left(\frac{\sigma_j}{\sigma_b}\right)^\frac{1}{\delta+1}} \right] \right \}^{\delta}. \nonumber
\end{eqnarray} 
The ratio between $t_j (j:c, a, b)$ and $t_h$  which  satisfies the following  relation,
\begin{eqnarray}\label{gtratio}
 \left(\frac{t_j}{t_h}\right)^{\frac{1}{\delta}+1} = \frac{\alpha_j T_c}{\alpha_h T_h}\left(\frac{\sigma_j}{\sigma_h}\right)^\frac{1}{\delta}.
\end{eqnarray}
Combining  Eqs.(\ref{geffg1}), (\ref{gtopt}) and (\ref{gtratio}), the efficiency at maximum power is obtained and is given below:
\begin{eqnarray}\label{geffmaxp}
   \eta_P	&=& \left(\frac{1}{\delta+1}\right)\frac{\eta_C}{1-\frac{\eta_C}{(1+\frac{1}{\delta})\zeta}},
\end{eqnarray}
where, 
\begin{eqnarray}\label{chie}
\zeta=1+\varsigma \left(\frac{T_c}{T_h}\right)^\frac{\delta}{\delta+1}=1+\varsigma \left(1-\eta_C\right)^\frac{\delta}{\delta+1} 
\end{eqnarray}
and 
\begin{eqnarray}\label{tun}
\varsigma&=&\sum_{j=c,a,b}{\left(\frac{\alpha_j}{\alpha_h}\right)^\frac{\delta}{\delta+1}\left(\frac{\sigma_j}{\sigma_h}\right)^\frac{1}{\delta+1}}.
\end{eqnarray}
The above equation shows that  $\eta_P$ in general does not exhibit any universal feature \cite{nonuniv}
and it depends only on the ratios of the individual parameters.

The power law dissipation of the phenomenological model with the non-adiabatic dissipation 
also governs the universal form of the efficiency under the assumption that the temperature difference 
between the two reservoirs is small, which is obtained by expanding $\eta_P$ in terms of $\eta_C$ as \cite{ponlow},
\begin{eqnarray}\label{uniiso}
  \eta_P=\left(\frac{1}{\delta+1}\right)\eta_C +\frac{\delta}{(\delta+1)^2 (1+\varsigma)}\eta_C^2+
	\frac{\delta^2}{(\delta+1)^3 (1+\varsigma)^2}\eta_C^3+.....
\end{eqnarray}
This shows that the non-adiabatic dissipation does not alter the super-universal feature 
of the efficiency at maximum power \cite{ponlow}. This is the first main result of the present paper.

\section{Discussions}

If we neglect the inner friction, the adiabatic dissipation coefficients, 
$\sigma_a \to 0$ and $\sigma_b \to 0$, Eqs.(\ref{geffmaxp}) and (\ref{uniiso}) reduces to 
the earlier results of power law dissipation Carnot like heat engines 
with instantaneous adiabatic processes \cite{ponlow}.
Now, we discuss whether the non-adiabatic dissipation will influence 
the universal nature  of the extreme bounds on the efficiency at maximum power for different 
cases of symmetric and asymmetric dissipation limits.  

It is observed from Eq.(\ref{geffmaxp}) that 
the value of $\eta_P$ depends mainly on the term $\varsigma$  (Eq.(\ref{tun})), 
which contains the ratios of dissipation coefficients and  tuning parameters \cite{ponlow}.
We obtain the minimum value of $\eta_P^-=\frac{1}{(\delta+1)}\eta_C$, when $\varsigma \to \infty$ 
and the maximum value of $\eta_P^+=\frac{\eta_C}{(\delta+1)-\delta \eta_C}$, when $\varsigma \to 0$.
This shows that $\eta_P$ lies between these two extreme bounds, which is given by,
\begin{eqnarray}\label{gbound}
\frac{1}{\delta+1}\eta_C \le \eta_P  \le  \frac{\eta_C}{(\delta+1)-\delta \eta_C}. 
\end{eqnarray}  
The lower and upper bounds can be obtained in the asymmetric dissipation limits of $\sigma_h \to 0$ and 
$\sigma_h \to \infty$, respectively,  for any finite values of $\sigma_j$ ($j:c,a,b$). 
Thus, one can obtain the generalized universal nature of lower and upper bounds on the efficiency at maximum power (Eq.(\ref{gbound})) under the combinations of isothermal and adiabatic asymmetric dissipation limits, which is same as the one obtained for power law dissipation Carnot-like heat engine model without having 
non-adiabatic disspation \cite{yang,ponlow}. This is the second main result of the preent paper.

In the case of completely symmetric dissipation, $\sigma_c=\sigma_h=\sigma_a=\sigma_b$,
the efficiency at maximum power (Eq.\ref{geffmaxp}) becomes,
\begin{eqnarray}\label{geffmaxpsym}
   \eta^{s}_P &=& \frac{\eta_C}{(\delta+1)-\frac{\delta\eta_C}{\zeta_{s}}},
\end{eqnarray}
where
\begin{eqnarray}\label{tunpsym} 
\zeta_{s}=1+\sum_{j=c,a,b}{\left(\frac{\alpha_j T_c}{\alpha_h T_h}\right)^\frac{\delta}{\delta+1}}.
\end{eqnarray}
Under the tuning condition, $\frac{\alpha_j}{\alpha_h}=\frac{1}{3}\frac{T_h}{T_c}$,  
$\eta^{s}_P$ reduces  to the efficiency at maximum power of the 
stochastic heat engine \cite{stocheng} with $\delta =1$.

\begin{figure}
\centering
\includegraphics[scale=0.4]{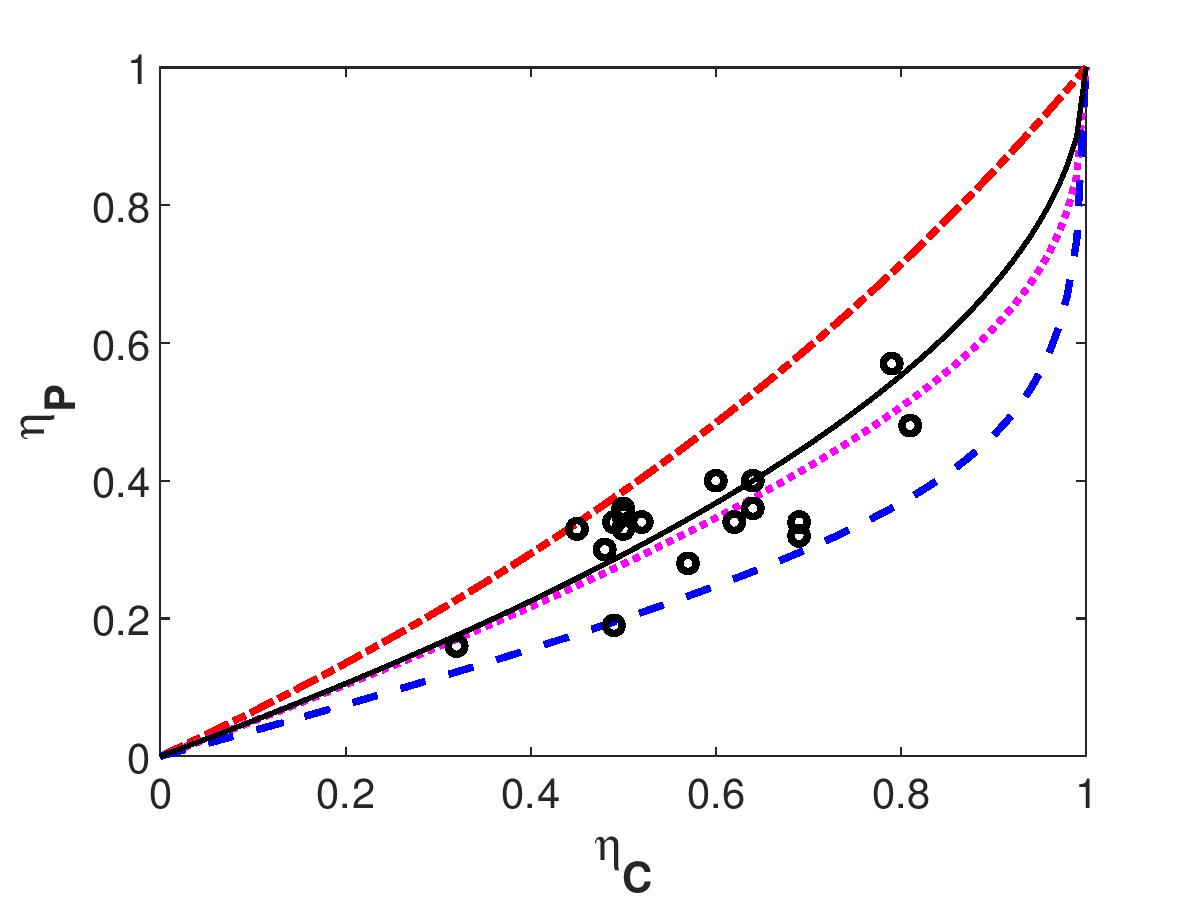}
\caption{
Efficiency at maximum power $\eta_P$ plotted as a function of  $\eta_C$
for different values of $\delta$ and $\varsigma$. Top (dot-dashed line): for 
$\delta=0.6$, $\varsigma=0$, Middle (dotted line): $\delta=1$ and $\varsigma=2$
and bottom (dashed line): $\delta=1.75$, $\varsigma=4$. The  observed efficiencies
of the various thermal power plants are shown in circles \cite{geneff,Espo,Johal}.
Solid line represents $\eta_{CA}$.
}\label{ebfig}
\end{figure}

Finally, in order to see the ranges of $\varsigma$ and $\delta$ in which $\eta_P$ covers 
the observed efficiencies of different thermal power plants \cite{Espo}, we plotted    
$\eta_P$ as a function of  $\eta_C$ in Figure.\ref{ebfig}
for different values of $\varsigma$ and $\delta$. 
The  observed efficiencies of the various thermal power plants are represented 
in circles \cite{geneff,Espo,Johal} and the solid line represents $\eta_{CA}$.
The figure shows $\eta_P$ encompasses $\eta_{CA}$ and the observed efficiencies of different power plants, 
which are in the combinations of isothermal and 
non-adiabatic dissipations range $0 \leq  \varsigma  \leq 4$ with the
power law dissipation range $0.6 \leq  \delta  \leq 1.75$. 

\section{Conclusion}
	We calculated the efficiency at the maximum power of the power law dissipation Carnot-like heat engines which 
taken in to account the non-adiabatic dissipation of finite time adiabatic processes. We found that the presence of non-adiabatic dissipation does not influence the universal feature of the generalized  extreme bounds on the  efficiency at maximum power. We expanded $\eta_P$ in terms of $\eta_C$ and also obtained the same 
universal form of the efficiency at  maximum power.


\end{document}